\documentstyle[aps,prl,twocolumn,psfig]{revtex}
\draft
\begin{document}
\title{Semiclassical Description of Tunneling in Mixed Systems:\\
 the Case of the Annular Billiard }
\author{E.\ Doron and S.\ D.\ Frischat}
\address{Max--Planck--Institut f\"ur Kernphysik, Postfach 103980,
69029 Heidelberg, F.R.G.}

\date{\today}

\twocolumn[
\maketitle
\widetext%
\vspace*{-6.45mm}
\leftskip=1.9cm\rightskip=1.9cm
\begin{abstract}
$\quad$We study quantum-mechanical tunneling between symmetry-related pairs
of regular phase space regions that are separated by a chaotic
layer. We consider the annular billiard, and use scattering theory to
relate the splitting of quasi-degenerate states quantized on
the two regular regions to specific paths connecting them.
The tunneling amplitudes involved are given a semiclassical
interpretation by extending the billiard boundaries to complex space
and generalizing specular reflection to complex rays. We give
analytical expressions for the splittings, and show that the
dominant contributions come from {\em chaos-assisted}\/ paths that
tunnel into and out of the chaotic layer.
%
\vspace*{-1mm}
\vspace*{-.1cm}
\end{abstract}
  ]

\narrowtext
Recently there has been a surge of interest in quantum systems whose
classical counterparts exhibit a mixture of regular and chaotic
motion. Such systems are of great importance since they comprise the
majority of dynamical systems found in nature. One interesting
feature is quantum-mechanical tunneling
between regular regions in phase space that are separated by a chaotic
layer.  Numerical studies of mixed systems have shown that
the splitting of quasi-doublets associated with such pairs of regular
regions was much higher than could be explained by direct tunneling
processes \cite{Lin90,BTU,Bohigas93a}.
This was attributed in \cite{BTU,Bohigas93a} to a suggested mechanism of {\em
chaos-assisted tunneling}, i.e.\ tunneling from one regular region
into the chaotic sea, propagating ``classically'' to the other side,
and tunneling out into the second regular region. Since a large part
of the phase space is thus traversed via classically allowed
transitions, these paths were expected to considerably enhance the
splitting.

In this work we study tunneling from a scattering
theory point of view, taking as a specific example the annular
billiard proposed in \cite{Bohigas93a}. This allows to write
the splittings as a sum over paths in angular momentum space, which
can include both classically allowed and tunneling transitions. The
tunneling amplitudes are evaluated semiclassically by continuing the
billiard boundaries to complex configuration space, and generalizing
the mechanism of specular reflection to the case of complex rays. We
obtain analytical expressions for the splittings in terms of these
transition amplitudes, and show that the dominant contributions arise
from paths that tunnel into (and out of) the chaotic layer via
intermediate angular momenta which lie on the boundary between the
regular and chaotic regions. This approach yields values for the
splitting which are in good agreement with exact results.

The annular billiard consists of
the space between two non-concentric circles of radius $R$ and $a<R$,
centered at $(x,y)$ coordinates $O\equiv(0,0)$ and $O'\equiv(-\delta,0)$,
respectively. Classically, a particle moves freely between
specular reflections on the bounding circles. We parametrize trajectories by
their impact parameter $L$ with respect to $O$ and their direction of
propagation $\gamma$. Trajectories of $|L|>a+\delta$ do not hit the inner
circle, but rotate forever at constant $L$, while the phase space for
$|L|<a+\delta$ consists of a mixture of regular islands and chaotic
layers. Phase space plots can be found in Ref.~\cite{Bohigas93a}.
In this work we use parameters for which a single chaotic layer extends
from $L=a+\delta$ to $L=-(a+\delta)$.

We apply the scattering approach to quantization first proposed in
\cite{Doron/Smil}. Let us write the wave function in
terms of incoming and outgoing cylindrical waves,
\begin{displaymath}
  \psi(r,\phi) = \sum_{n=-\infty}^{\infty}\left[
    \alpha_n \mathop{\rm H{}}^{\text{\,(2)}}\nolimits_n(kr) + \beta_n
      \mathop{\rm H{}}^{\,(1)}\nolimits_n(kr)\,\right] \text{e}^{in\phi} \:,
\end{displaymath}
where $\mathop{\rm H{}}^{\,(1,2)}\nolimits_n(x)$ denotes the Hankel
functions of first and second kind, and $k$ is the wave number. We
consider the billiard as two back-to-back scattering systems; the
first, ``inner'' system consists of incoming waves being reflected to
outgoing waves by the exterior of the inner circle, whereas the
``outer'' system consists of outgoing waves being scattered to
incoming waves by the interior of the outer circle. The two system are
characterized by scattering matrices $S^{\,\text{(I,O)}}(k)$,
respectively, that relate the coefficient vectors $\bbox{\alpha}\,$,
$\bbox{\beta}\:$ by
\mbox{$\bbox{\beta}=S^{\,\text{(I)}}(k)\,\bbox{\alpha}\:$} and
\mbox{$\bbox{\alpha}=S^{\,\text{(O)}}(k)\,\bbox{\beta}\:$}. Requiring the two
relations to be consistent results in the quantization condition
\begin{equation}
  \det(S(k)-1) = 0\:, \quad
  S(k)=S^{\,\text{(I)}}(k) S^{\,\text{(O)}}(k)\:.
\label{quantize}\end{equation}
Thus the billiard supports an eigenvalue whenever one of the eigenphases of
$S(k)$ equals an integer multiple of $2\pi$.

We drop the wave number argument in the sequel.  $S^{\,\text{(O)}}$ is
clearly just the diagonal matrix $S^{\,\text{(O)}}_{nm}=-\mathop{\rm
H{}}^{\,(1)}\nolimits_n(kR)/\mathop{\rm
H{}}^{\text{\,(2)}}\nolimits_n(kR)\:\delta_{nm}$, while
$S^{\,\text{(I)}}$ can be obtained by writing it as
\mbox{$S^{\,\prime\,{\text{(I)}}}_{nm}=-\mathop{\rm
H{}}^{\text{\,(2)}}\nolimits_n(ka)/\mathop{\rm
H{}}^{\,(1)}\nolimits_n(ka)\:\delta_{nm}$} in the primed coordinates,
and then performing a coordinate change. Using the addition theorem
for Bessel functions (see e.g.~\cite{Abramowitz}) this gives
\begin{equation}
  S^{\,\text{(I)}}_{nm} = -\sum_{\ell=-\infty}^\infty
  \mathop{\rm J{}}\nolimits_{n-\ell}(k\delta)\,
    \mathop{\rm J{}}\nolimits_{m-\ell}(k\delta)\:
  \frac{\mathop{\rm H{}}^{\text{\,(2)}}\nolimits_\ell(ka)}
    {\mathop{\rm H{}}^{\,(1)}\nolimits_\ell(ka)} \;.  \label{Sinternal}
\end{equation}

The structure of $S$ can be seen in Fig.~\ref{fig1}, which shows
$|S_{nm}|$ for $k=100$, $a=0.4$ and $\delta=0.2$. The inset gives an
overview of one quadrant. One can see that $S$ is almost diagonal at
angular momenta $n,m>k(a+\delta)$. For $n,m$ below this value most of
the amplitude lies within the region of classically allowed
transitions and is delimited by caustics due to classical rainbow
scattering. Closer inspection reveals that these ridges extend into
the classically forbidden region as well, although they are
exponentially suppressed there. The main figure shows a single row
$n=70$. Since $n>k(a+\delta)$, the diagonal element is almost unimodular, while
the non-diagonal elements are exponentially small. One finds a maximum
at $m\approx 63$ that corresponds to the ridge seen in the inset,
while away from this maximum the matrix elements decay faster than
exponentially.

For our purpose it will prove sufficient to approximate the magnitudes
$|S_{nm}|=|S^{\,\text{(I)}}_{nm}|$.  We will now sketch this semiclassical
derivation; a more detailed account will be given elsewhere \cite{us_later}.
Starting from (\ref{Sinternal}), we write $\mathop{\rm
J{}}\nolimits_{m-\ell}(k\delta)=\text{e}^{i\pi(m-\ell)}\mathop{\rm
J{}}\nolimits_{\ell-m}(k\delta)$  and apply the Poisson summation
formula to get
\begin{eqnarray}
  S^{\,\text{(I)}}_{nm} =
  &-&\sum_{\mu=-\infty}^\infty \int_{-\infty}^\infty \text{d}\ell\;
        \text{e}^{i2\pi \mu\ell+i\pi(m-\ell)}
\nonumber\\
  &\times&  \mathop{\rm J{}}\nolimits_{n-\ell} (k\delta)
      \mathop{\rm J{}}\nolimits_{\ell-m}(k\delta)
     \:\frac{\mathop{\rm H{}}^{\text{\,(2)}}\nolimits_\ell(ka)}
        {\mathop{\rm H{}}^{\,(1)}\nolimits_\ell(ka)}\;,
   \label{poisson}
\end{eqnarray}
and consider only the $\mu=0$ term (see below).
We replace the Bessel functions by their Sommerfeld representation
$ \mathop{\rm J{}}\nolimits_\nu(z) = 1/ (2\pi)\int_C d\gamma\,
    \exp[iz\cos\gamma+i\nu\left(\gamma-{\pi/ 2}\right)]$,
where the contour of integration $C$ runs from $-\pi+i\infty$ to
$\pi+i\infty$. The contribution of the Hankel functions to non-diagonal
elements of $S^{\,\text{(I)}}$ is asymptotically given by their Debye
approximation \cite{Sommerf} to be $i\exp[i\,2ka(\sin Q-Q\cos Q)]$, where
$Q=\arccos(\ell/ka)$.
$S^{\,\text{(I)}}_{nm}$ can then be written in the form of an
integral over an exponential whose argument is proportional to $k$. In the
semiclassical limit $k\gg 1$ this integration can be carried out using the
saddle point approximation, giving
\begin{eqnarray}
  && S^{\,\text{(I)}}_{nm} \approx \sum_p
    \sqrt{\frac{|{\cal R}_p|}{k}} \;
    \text{e}^{i k \Phi_p + {i\over 2}\arg{\cal R}_p -i{3\pi\over 4 }}
    \;, \nonumber\\
  \Phi_p &=& \delta(\cos\gamma_{f,p}-\cos\gamma_{i,p}) -
       {m\over k}\left(\gamma_{f,p}+{\pi\over 2}\right)
\nonumber\\
        &&+
       {n\over k}\left(\gamma_{i,p}+{\pi\over 2}\right) +
       2a\sin\left({\gamma_{f,p}-\gamma_{i,p} \over 2}\right)\;, \nonumber\\
   {\cal R}_p &=& \frac{1}{2\pi}\:\frac{\partial^{\,2} \Phi}
             {\partial\gamma_f\partial\gamma_i}
      \left[
        \frac{\partial^{\,2}\Phi}{\partial\gamma_f^2}\,
        \frac{\partial^{\,2}\Phi}
             {\partial\gamma_i^2}
        -\left(\frac{\partial^{\,2} \Phi}
                    {\partial\gamma_f\partial\gamma_i}  \right)^2
     \right]^{-1}_{p}\;,
    \label{sadcont}
\end{eqnarray}
where the summation is over saddles $p$ in the complex $Q,\gamma_i,\gamma_f$
space. These saddles are determined by
\begin{mathletters}\label{saddles}
\begin{eqnarray}
  0&=&\gamma_{f,p} - \gamma_{i,p}\:+2Q_p  \;,\label{specular}\\
  m &=& n - k\delta\,(\sin\gamma_{f,p}-\sin\gamma_{i,p})\;,
        \label{deflect}\\
  n &=& -k\delta\sin\gamma_{i,p}+ka\cos Q_p\;. \label{transf}
\end{eqnarray}
\end{mathletters}
Solving (\ref{saddles}) yields a number of saddle points in the complex
plane. For most values of $n,m$ the dominant contribution comes from a single
saddle, whereas others are either negligible or cannot be reached by
deformation of the contour of integration. This dominant saddle coalesces
with its symmetry-related counterpart at values of $n,m$ which lie close to
the ridges seen in the inset of Fig.~\ref{fig1}. (The actual ridges
correspond to points where $\gamma_{f,p}$ becomes real and equal to $\pi/2$
or $3\pi/2$). In principle, near this caustic saddle point integration should
be replaced by a uniform approximation, and also the Hankel functions should
be treated more carefully. However, a detailed treatment of this region is
not required for the purpose of this paper. The full line in
Fig.~\ref{fig1} shows the contribution of the dominant saddle to
$|S^{\,\text{(I)}}_{nm}|$, and one can see that there is excellent agreement
over a wide range of angular momenta.

For completeness, let us consider the $\mu\ne 0$ terms in
(\ref{poisson}). Each integral in the summation can be written as a
residue sum over the poles of $\mathop{\rm
H{}}^{\text{\,(2)}}\nolimits_\ell(ka)/\mathop{\rm
H{}}^{\,(1)}\nolimits_\ell(ka)$. These residue contributions are interpreted
in standard diffraction theory as surface waves excited on the
inner disc by rays of grazing incidence \cite{Nussenz}.
As can be seen from Fig.~\ref{fig1}, at our parameter values they
give rise to significant corrections only
over a small range of $m$. As the corresponding transitions will turn
out to be of no interest in the present context, we defer a
discussion of the residue terms to \cite{us_later}.

The approximation given in Eqs.\ (\ref{sadcont}), (\ref{saddles}) can
be interpreted in terms of classical trajectories in complex
configuration space, by extending the dynamics inside the billiard to
complex coordinates.  The extension of free flight to complex
coordinates is trivial. In order to describe the interaction with the
inner circle, we define a complex circle of radius $a$ around the
point $(x_0,y_0)$ as the surface for which $x-x_0=a\cos\beta$,
$y-y_0=a\sin\beta$, where $\beta$ is allowed to be complex. Our
motivation is that a wave function satisfying Dirichlet or Neumann
boundary conditions on such a circle for real $\beta$ will also
satisfy them for complex $\beta$. Next, we note that through (almost)
every point on a circle one can pass two distinct trajectories with a
given impact parameter relative to its center. Since reflection from a
circle centered at the origin preserves the impact parameter, we
define specular reflection as the mapping from one of these trajectories
to the other.

It is easy to verify that any given initial and final impact
parameters are linked by at least one complex trajectory. The
conditions specifying the corresponding initial and final angles are
then equivalent to Eqs.~(\ref{saddles}). Thus $\gamma_{i,p}$ and
$\gamma_{f,p}$ have the meaning of the initial and final angles for
which an incoming trajectory with impact parameter $L_i=n/k$ is
reflected off the complex inner circle to the impact parameter
$L_f=m/k$. Moreover, (\ref{specular}) is the generalized specular
reflection condition, while (\ref{deflect},\ref{transf}) give the
classical deflection function. Note that the impact parameter with
respect to the center of the inner circle is given by $L' = a\cos
Q$. By (\ref{transf}), we see that $L'$ can become complex even if
$L_i$ and $L_f$ are real. Furthermore, unlike in \cite{Shudo95}, both
the initial {\em and\/} final angles are generically complex.

Finally, it is straightforward to show that the phase $\Phi$ in
(\ref{sadcont}) is given by the reduced action ({\em not\/} the length) of
the ray,
\mbox{$\Phi = -\int_{t_i}^{t_f} \text{d} t\:[ r(t)\dot{p}_r(t)+
      \varphi(t)\dot{L}(t)]$},
where $(r,\varphi)$, $(p_r,L)$ are canonically conjugate polar coordinates
and momenta, and that $|{\cal R}_p|$ is the corresponding reciprocal stability
$|\partial^{\,2}\Phi/\partial L_i \partial L_f|/2\pi$.
We therefore find that (\ref{sadcont}) coincides
with the sum over classical trajectories which constitutes the usual
semiclassical approximation to scattering matrix elements (see \cite{scatt}).

We will now extract the level splitting. We first examine the
splittings in eigenphases, and will regard energy splittings later. Let
$|\,\pm\,\rangle$ represent the doublet of eigenvectors peaked at
angular momenta $\pm n$,
\begin{displaymath}
|\pm\:\rangle =\frac{1}{\sqrt{2}}\:\Big(\,\mid n\:\rangle \pm \mid -n
\:\rangle\,\Big)
+ \sum_m\kappa^\pm_m \mid m \:\rangle\;,
\end{displaymath}
where the $\kappa^\pm_m$ are exponentially small.
We denote the corresponding eigenphases by $\theta^{\pm}_n$ and their
splitting by $\delta\theta_n=|\theta_n^+ -\theta_n^-|$.
Using \mbox{$\exp(iN\theta_n^\pm)=\langle\,\pm\,| S^N|\,\pm\,\rangle$} we get
\begin{displaymath}
\sin\left(\frac{N}{2}\,\delta\theta_n\right)
=\left| [S^N]_{-n,n}\right| + {\cal C}^{(N)}_n\;,
\end{displaymath}
which describes the tunneling oscillations between $n$ and
$-n$, with a correction term ${\cal C}^{(N)}_n \lesssim
k^2\max|\kappa_m^\pm|^2$. For ${\cal C}^{(N)}_n\ll N\,\delta\theta_n
\ll 1$, we can therefore write
\begin{equation}
 \delta\theta_n \approx\frac{2}{N}\:\left| [ S^N ]_{-n,n}\right|
= \frac{2}{N}\:\Big|\sum_{\{\lambda_i\}}
  \prod_{i=1}^{N-1} \:S_{\lambda_i,\lambda_{i+1}}\Big|\:.
\label{Split}\end{equation}

Eq.~(\ref{Split}) relates the splitting to a sum over all paths
$\{\lambda_i\}_{i=1}^N$ in matrix element space that go from
$\lambda_1=-n$ to $\lambda_N=n$.  The simplest such paths involve only
the direct tunneling amplitude $S_{-n,n}$, which from Fig.~\ref{fig1}
is exceedingly small.  A much larger contribution comes from paths
which tunnel only over small distances in angular momentum space and
traverse the remaining distance via classically allowed transitions.
These are the chaos-assisted tunneling paths proposed in
\cite{BTU}. In order to deal with such contributions, we first need to
consider the structure of the chaotic part of phase space.

The quantum evolution of systems with purely chaotic classical
counterparts is usually very well described by random matrix theory
\cite{RMT}. However, this is only appropriate when phase
space can be assumed to be completely structureless. In the case of a
classically mixed system, this is an over-simplification: as the
Lyapunov exponent vanishes smoothly at the interface between chaotic
and regular phase space regions, there is always an intermediate layer
in which chaotic, but relatively stable motion gives rise to classical
staying times much longer than the mean level
density. Moreover, angular momentum remains a preferred basis well
into the chaotic part of phase space. This corresponds to the
observation made in \cite{BTU,Utermann94} that classical regular
structure can be quantum mechanically continued into the chaotic sea.
Consequently, it is only the internal part $|\,l\,|\le l_{\text{COE}}$ of
the chaotic sea that is appropriately modelled by a random matrix
ensemble, which in the present case is a circular orthogonal ensemble
(COE) of size $\sim 2l_{\text{COE}}$.

It turns out that there are two types of chaos-assisted paths that
contribute dominantly to the splitting: (I) paths $(n,\gamma,-n)$ that
tunnel from $n$ directly into the chaotic region, propagate in some
eigenstate $|\gamma\rangle$ of the internal block, and finally tunnel
to $-n$, and (II) paths $(n,l,\gamma,-l,-n)$ that propagate from $n$
to $|\,\gamma\,\rangle$ and then to $-n$ in two jumps via intermediate
edge angular momenta $l$ and $-l$, respectively.  We will first
consider paths of type (I). Let us denote the eigenphase of
$|\,\gamma\,\rangle$ by $\theta_\gamma$, and let $\theta_n=-i\log
S_{n,n}$. Summing over the possible dwell times at $n$ and $\gamma$
gives contributions of the type \mbox{$S_{n,\gamma}\,S_{\gamma,-n}/
\sin[(\theta_\gamma -\theta_n)/2]$} to the splitting.  Assuming that
the $S_{n,\gamma}$ are distributed independently of the
$\theta_\gamma$, we can average over the COE and calculate the median
of the type (I) splitting contributions \cite{Felix}
\begin{equation}
\delta\theta_n^{(I)} = \left|\sum_\gamma
\frac{S_{n,\gamma}\,S_{\gamma,-n}}{\sin[(\theta_\gamma -\theta_n)/2]}
\right| \sim \frac{8}{\pi} \:v_n^2\ ,
\label{typeone}\end{equation}
where $v_n^2=\langle\, \sum_\gamma | S_{n,\gamma}|^2\,\rangle$,
which we approximate by \mbox{$v^2_n \approx \sum_{g=- l_{\text{COE}}}^{
l_{\text{COE}}} |\,S_{n,g}\,|^2$}.  Similarly, the type (II)
contributions can be estimated by
\begin{eqnarray}
\delta\theta_n^{(II)} &=& \frac{1}{4}\left| \sum_{l,\gamma}
\frac{S_{n,l}\,S_{-l,-n}}{\sin^2[(\theta_l-\theta_n)/2]}\:
\frac{S_{l,\gamma}\,S_{\gamma,-l}}{\sin[(\theta_\gamma -\theta_l)/2]}
\right| \nonumber\\
&\sim& \frac{2}{\pi} \:\left(\,{\sum_{l}}
\:\left|\: \frac{\mid S_{n,l}\mid^2\: v^2_l}
{\sin^2[(\theta_l-\theta_n)/2]} \:\right|^2 \:\right)^{1/2}\ .
\label{typetwo}\end{eqnarray}
Note that $\delta\theta_n$ will fluctuate strongly about the
median values due to avoided crossings
\cite{BTU,Bohigas93a,Leyvraz95}.  The relative importance of the contributions
$\delta\theta_n^{(I,II)}$ is system-specific; for the present system
we find $\delta \theta_n^{(II)}/\delta\theta_n^{(I)} \sim
25$. Furthermore, paths that include additional transitions outside
the chaotic block give negligible contributions.

It remains to connect eigenphase and eigenvalue splittings. From
(\ref{typeone},\ref{typetwo}) we see that $\delta\theta_n(k)$ is
approximately constant over exponentially small ranges of $k$. It is
then simple to show that \mbox{$\delta k_n \approx
|\,\partial\theta_n^{\,(0)}\!/\partial k\,|^{-1} \delta\theta_n$}, where
$\theta_n^{\,(0)}$ is evaluated at $\delta = 0$.
However, we expect Eqs.~(\ref{typeone},\ref{typetwo}) to
over-estimate the actual splittings: our model neglects residual
transport barriers inside the chaotic block, and the damping effect of
imaginary parts of eigenphases is not accounted for. As these errors
are expected to be independent of $n$, we can correct by an overall
factor $c$ that we extract from the numerical data.

In Fig.~\ref{fig2} we present a numerical test of our results. We
plot median values of $\delta\theta_n$ that were obtained
by varying the outer radius $R$ over $30$ values between
$1$ and $1.3$. Note that changing $R$ leaves
transition probabilities constant.  The full circles represent the exact
$\delta\theta_n$, as function of $n$. Splittings smaller than
$\sim~10^{-14}$ (denoted by empty circles) could not be calculated
directly due to the finite precision of the diagonalization procedure,
but were obtained by calculating $S^N$ for large $N$ and applying
Eq.~(\ref{Split}). Finally, the dashed line shows the prediction of
Eqs.~(\ref{typeone},\ref{typetwo}) for $l_{\text{COE}}=50$ and
$c=0.1$. While the $n$-dependence due to tunneling is reproduced very
well, the coefficient due to transport across the chaotic sea
could only be calculated up to an order of magnitude.
The inset shows $\delta k_n/\delta \theta_n$ evaluated from
the quantization condition (\ref{quantize}) around $k= 100$ for $R=1$,
compared to $|\,\partial\theta_n^{\,(0)}\!/\partial
k\,|^{-1}$. We see that the correspondence is very good.

It is a pleasure to thank H.\ A.\ Weidenm\"uller for pointing out this
problem to us and for fruitful discussions.
After submission of this Letter we received a preprint \cite{takada95}
that deals with related topics.
This work was partially funded by a grant from MINERVA.

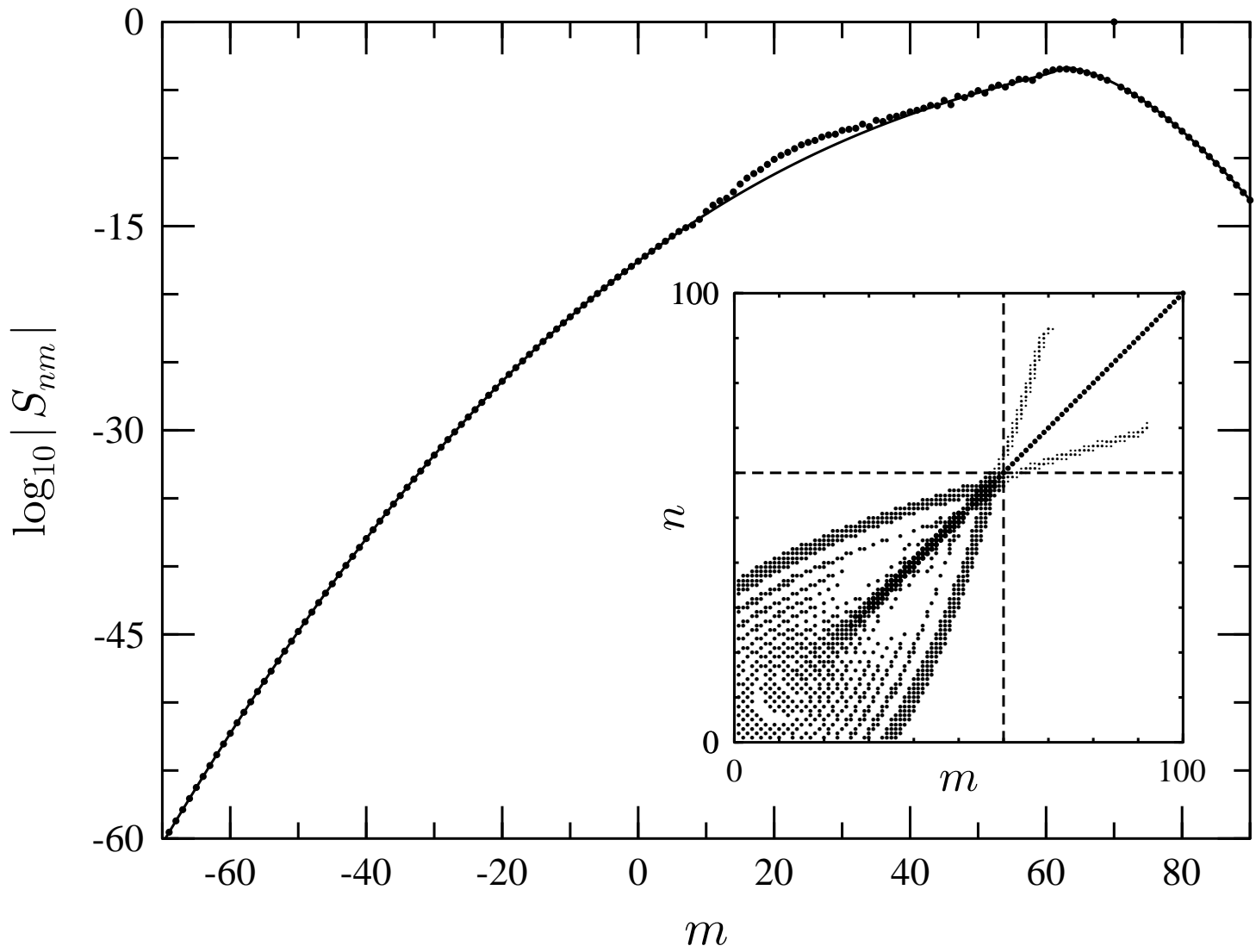
\begin{figure}
\caption{ Tunneling amplitudes $|S_{nm}|$ for $n=70$, $a=0.4$,
  $\delta=0.2$ and $k=100$ as a function of $m$, calculated exactly
  (dots) and semiclassically (full line). The inset shows one
  quadrant of $|S|$, with larger values corresponding to larger
  dots, in arbitrary units. Off-diagonal ridges at $n,m > k(a+\delta)$
  are exponentially enhanced in the plot. The dashed lines indicate
  $k(a+\delta)$.
\label{fig1}}
\end{figure}
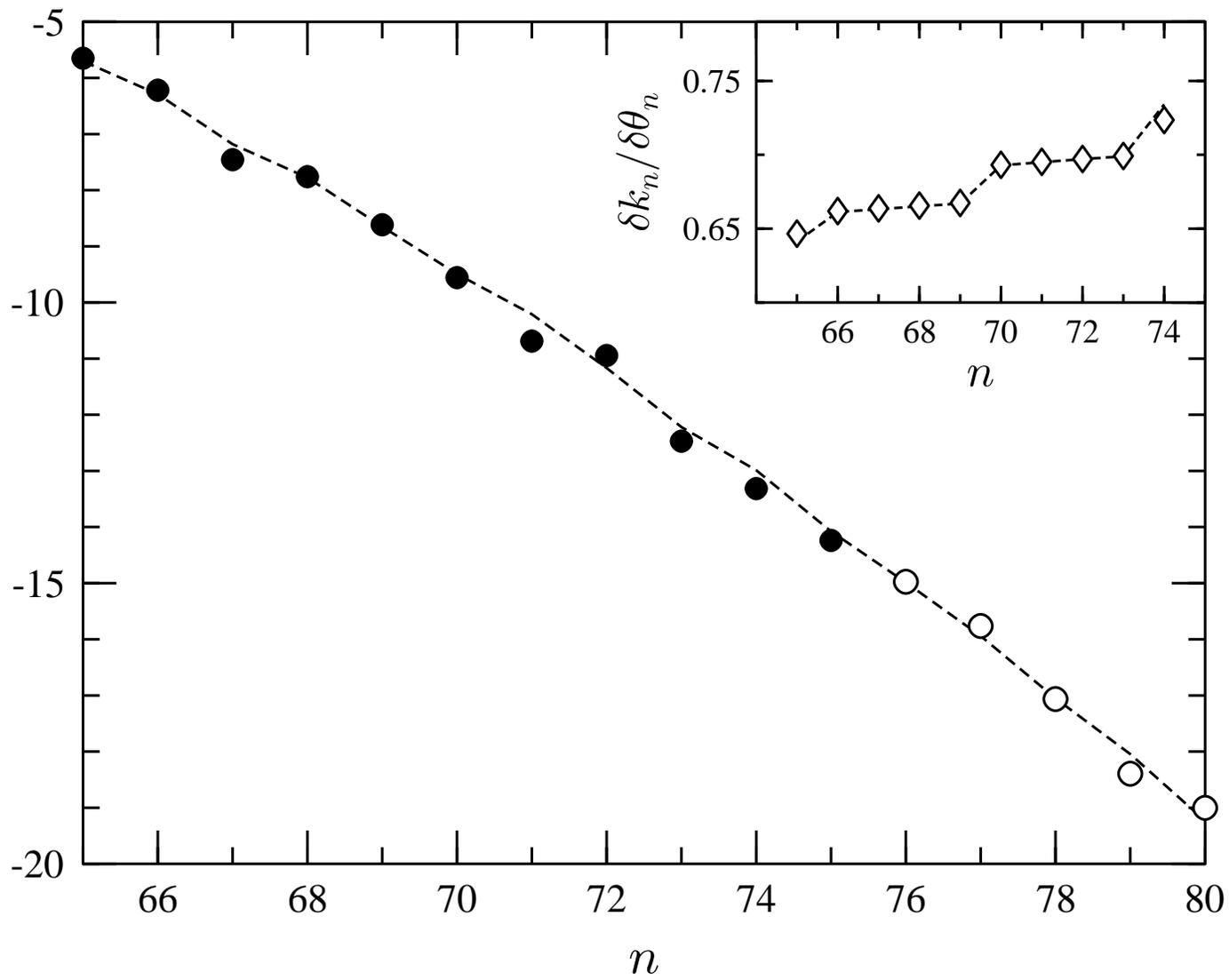
\begin{figure}
\caption{Median eigenphase splittings $\delta\theta_n$ obtained by
  diagonalization of $S$ (full circles), using
  Eq.~(\protect\ref{Split}) with $N=2^{13}$ (empty circles), and as
  estimated from Eqs.~(\protect\ref{typeone},\protect\ref{typetwo})
  (dashed line) with $c=0.1$ (see text). Inset: $\delta
  k_n/\delta\theta_n$ (diamonds) and
  $|\,\partial\theta_n^{(0)}\!/\partial k\,|^{-1}$ (dashed line) evaluated
  around $k= 100$.
\label{fig2} }
\end{figure}


\begin{references}
\bibitem{Lin90} W. A. Lin and L. E. Ballentine, Phys. Rev. Lett. {\bf
    24} 2927 (1990).
\bibitem{BTU}O. Bohigas, S. Tomsovic and D. Ullmo, Phys. Rep. {\bf
    223} 43 (1993); S. Tomsovic and D. Ullmo, Phys. Rev. E {\bf 50}
  145 (1994).
\bibitem{Bohigas93a}O. Bohigas, D. Boos\'e, R. Egydio de Carvalho and
  V. Marvulle, Nucl. Phys. A {\bf 560} 197 (1993).
\bibitem{Doron/Smil}E. Doron and U. Smilansky, Phys. Rev. Lett. {\bf
    68} 1255 (1992); E. Doron and U. Smilansky, Nonlinearity {\bf 5}
  1055 (1992).
\bibitem{Abramowitz}{\em Handbook of Mathematical Functions}, edited
  by M. Abramowitz and I. A. Stegun, (Dover, New York, 1972).
\bibitem{us_later}E. Doron and S. D. Frischat, unpublished.
\bibitem{Sommerf}A. Sommerfeld, {\em Lectures on Theoretical Physics},
  Vol. 6 (Academic Press, 1964).
\bibitem{Nussenz}H. M. Nussenzveig, Ann. Phys. (NY) {\bf 34} 23 (1965).
\bibitem{Shudo95}A. Shudo and K. S. Ikeda, Phys. Rev. Lett. {\bf 74}
  682 (1995).
\bibitem{scatt}R. A. Marcus, J. Chem. Phys. {\bf 54} 3965 (1971);
  W. H. Miller, Adv. Chem. Phys. {\bf 25} 69 (1972); U. Smilansky, in
  {\em Chaos and Quantum Physics}, ed. by M.-J. Giannoni, A. Voros
  and J. Zinn-Justin (Elsevier, Amsterdam, 1992).
\bibitem{RMT}C. E. Porter, {\em Statistical Theories of Spectral
    Fluctuations} (Academic Press, 1965).
\bibitem{Utermann94}R. Utermann, T. Dittrich and P. H\"anggi,
  Phys. Rev. A {\bf 49} 273 (1994).
\bibitem{Felix}F. von Oppen, private communication.
\bibitem{Leyvraz95}F. Leyvraz and D. Ullmo,
  preprint (1995).
\bibitem{takada95}S. Takada, P. N. Walker and M. Wilkinson, preprint
  (1995).

\end{references}
\end{document}